\documentclass[epj]{webofc}
\usepackage[utf8]{inputenc}
\usepackage[varg]{txfonts}   
\usepackage{booktabs}
\usepackage{xcolor}
\definecolor{darkred}{rgb}{0.4,0.0,0.0}
\definecolor{darkgreen}{rgb}{0.0,0.4,0.0}
\definecolor{darkblue}{rgb}{0.0,0.0,0.4}
\usepackage[bookmarks,linktocpage,colorlinks,
    linkcolor = darkred,
    urlcolor  = darkblue,
    citecolor = darkgreen]{hyperref}
\wocname{EPJ Web of Conferences}
\woctitle{Lattice2017}
\newcommand{\Slash}[1]{{\ooalign{\hfil/\hfil\crcr$#1$}}}
\begin{document}
%
\selectlanguage{english}
\title{4D $\mathcal{N}=1$ \,SYM supercurrent on the lattice in terms of the gradient flow
}
\author{%
\firstname{Kenji} \lastname{Hieda}\inst{1} \and
\firstname{Aya} \lastname{Kasai}\inst{1}\fnsep\thanks{Speaker, %
\email{kasai@phys.kyushu-u.ac.jp}} \and
\firstname{Hiroki}  \lastname{Makino}\inst{1} \and
\firstname{Hiroshi} \lastname{Suzuki}\inst{1}
}
\institute{%
Department of Physics, Kyushu University, 744 Motooka, Nishi-ku, Fukuoka, 819-0395, Japan
}
\abstract{%
The gradient flow~\cite{Narayanan:2006rf,Luscher:2009eq,Luscher:2010iy,%
Luscher:2011bx,Luscher:2013cpa} gives rise to a versatile method to construct
renormalized composite operators in a regularization-independent manner. By
adopting this method, the authors
of~Refs.~\cite{Suzuki:2013gza,Makino:2014taa,Endo:2015iea,Hieda:2016lly}
obtained the expression of Noether currents on the lattice in the cases where
the associated symmetries are broken by lattice regularization. We apply the
same method to the Noether current associated with supersymmetry, i.e., the
supercurrent. We consider the 4D $\mathcal{N}=1$ super Yang--Mills theory and
calculate the renormalized supercurrent in the one-loop level in the
Wess--Zumino gauge. We then re-express this supercurrent in terms of the flowed
gauge and flowed gaugino fields~\cite{Hieda:2017sqq}.
}
\maketitle
\section{Introduction}
\label{intro}

Lattice gauge theory provides a non-perturbative definition of quantum field
theory (QFT) and a powerful tool of simulating it. In this framework, QFT is
regularized by discretizing the spacetime and hence the continuum spacetime
symmetries are explicitly broken. Although these symmetries are often expected
to be restored in the continuum limit, this fact complicates the construction
of the Noether current associated with those spacetime symmetries, e.g.
the energy-momentum tensor.

In general, composite operators such as Noether currents potentially have UV
divergences and hence need regularization. In order to construct the composite
operators in a regularization-independent manner, we consider the gradient
flow which is defined for the gauge field
by
\begin{equation}
   \partial_{t}B_\mu(t,x)
   =D_\nu G_{\nu\mu}(t,x),\qquad
   B_\mu(t=0,x)=A_\mu(x),
\label{eq:(1)}
\end{equation}
where
\begin{equation}
   G_{\mu\nu}(t,x)
   =\partial_\mu B_\nu(t,x)-\partial_\nu B_\mu(t,x)
   +\left[B_\mu(t,x),B_\nu(t,x)\right],
   \qquad
   D_\mu=\partial_\mu+\left[B_\mu,\cdot\right].
\label{eq:(2)}
\end{equation}
It was proved in~Ref.~\cite{Luscher:2011bx} that composite operators composed
of the flowed gauge field are UV finite and therefore independent of
regularization~(see also Ref.~\cite{Hieda:2016xpq}). Also, a small flow-time
limit~$t\to0$ of a bare composite operator at~$t>0$ can be expanded by
renormalized composite operators at~$t=0$~\cite{Luscher:2011bx}. By using
these facts, one can express for example the chiral condensate
$\langle\Bar{q}(x)q(x)\rangle$ in massless QCD in terms of a $t\to0$ limit
of a composite operator at~$t>0$~\cite{Endo:2015iea,Hieda:2016lly}. Also, the
renormalized energy--momentum tensor can be expressed in terms of flowed bare
composite operators~\cite{Suzuki:2013gza,Makino:2014taa}. The validity of these
representations has been numerically confirmed
in~Refs.~\cite{Asakawa:2013laa,Taniguchi:2016ofw,Kitazawa:2016dsl,%
Taniguchi:2016tjc}.

In the present study, we apply the same method to construct the supercurrent in
the 4D $\mathcal{N}=1$ super Yang--Mills theory (SYM).%
\footnote{In~Ref.~\cite{Kikuchi:2014rla}, the gradient flow in this system is
studied from a quite different perspective.} Supersymmetry can be a
crucial element in theories beyond the standard model, providing a solution to
the fine-tuning problem in the Higgs mass. It is thus interesting to explore
the prediction of supersymmetric models by using the lattice simulation.
However, since supersymmetry is a spacetime symmetry, it is explicitly broken
by lattice regularization and, as a consequence, one has to tune the bare
parameters towards the supersymmetric point. For this tuning, a priori
knowledge on the correct supercurrent which restores the conservation law in
the continuum limit will be quite helpful.\footnote{In the 4D $\mathcal{N}=1$
SYM, one may employ the chiral symmetry to carry out the tuning of the gaugino
mass~\cite{Curci:1986sm,Kaplan:1983sk}. See
also~Ref.~\cite{Ali:2016zke}.\label{foot-1}}

\section{Renormalized supercurrent in 4D $\mathcal{N}=1$ SYM}
\label{sec-1}
In order to construct a Noether current using the gradient flow, we proceed in
two steps. The first is a construction of the Noether current in a regularized
theory; the second is to express it by flowed bare composite operators. In
this section, we carry out the first step. That is, we find the expression of
the correctly-normalized supercurrent in the one-loop level by using
dimensional regularization.

The Euclidean action of the 4D $\mathcal{N}=1$ SYM is given by
\begin{equation}
   S=\frac{1}{4g_0^2}
   \int d^Dx\,F_{\mu\nu}^a(x)F_{\mu\nu}^a(x)
   +\frac{1}{2}\int d^Dx \,\Bar{\psi}^a(x)\Slash{D}^{ab}\psi^b(x).
\label{eq:(3)}
\end{equation}
We adopt dimensional regularization with~$D=4-2\epsilon$. The
gaugino~$\psi^a(x)$ is a Majorana fermion in the adjoint representation
satisfying $\Bar{\psi}(x)=\psi^T(x)(-C^{-1})$, where $C$ is the charge-conjugation matrix 
such that $C^{-1}\gamma_\mu C=-\gamma_\mu^T$. The gauge-field
strength~$F_{\mu\nu}(x)$ and the covariant derivative~$D_\mu$ are defined by
\begin{align}
   F_{\mu\nu}^a(x)
   &=\partial_\mu A_\nu^a(x)-\partial_\nu A_\mu^a(x)
   +f^{abc}A_\mu^b(x)A_\nu^c(x),
\label{eq:(4)}
\\
   D_\mu^{ab}&=\delta^{ab}\partial_\mu+f^{acb}A_\mu^c.
\label{eq:(5)}
\end{align}
The super transformation in the Wess--Zumino gauge is given by
\begin{equation}
   \delta_\xi A_\mu^a(x)
   =g_0\Bar{\xi}\gamma_\mu\psi^a(x),
   \qquad
   \delta_\xi\psi^a(x)
   =-\frac{1}{2g_0}\sigma_{\mu\nu}\xi F_{\mu\nu}^a(x),
  \qquad
  \delta_\xi\Bar{\psi}^a(x)
   =\frac{1}{2g_0}\Bar{\xi}\sigma_{\mu\nu}F_{\mu\nu}^a(x),
\label{eq:(6)}
\end{equation}
where $\sigma_{\mu\nu}=\frac{1}{2}[\gamma_\mu,\gamma_\nu]$. This transformation
leaves the action~$S$ invariant.\footnote{In order to prove the invariance,
one has to use the Fierz identity,
\begin{equation}
   (\Bar{\psi}_1\gamma_\mu\psi_2)
   (\Bar{\psi}_3\gamma_\mu\psi_4)
   =(\Bar{\psi}_1\gamma_\mu\psi_4)
   (\Bar{\psi}_3\gamma_\mu\psi_2).
\label{eq:(7)}
\end{equation}
If we use this relation, the variation of the action vanishes: 
\begin{equation}
   \delta_\xi S
   =-\frac{1}{2}g_0\int d^Dx\,
   f^{abc}\Bar{\xi}\gamma_\mu\psi^a(x)
   \Bar{\psi}^b(x)\gamma_\mu\psi^c(x)=0.
\label{eq:(8)}
\end{equation}
However, the Fierz identity is broken with dimensional regularization and
$\delta_\xi S$ cannot be neglected in quantum level; we have to take this
effect into account in the Ward--Takahashi relation.
\label{foot-2}}
We can read off the classical form of the supercurrent~$s_\mu(x)$ by making
the parameter local $\xi\to\xi(x)$ (the Noether method). The result is
\begin{equation}
   s_\mu(x)
   =-\frac{1}{2g_0}\sigma_{\rho\sigma}\gamma_\mu\psi^a(x)F_{\rho\sigma}^a(x).
\label{eq:(9)}
\end{equation}

We next consider the Ward--Takahashi (WT) relation associated with
supersymmetry and find the expression of the correctly-normalized supercurrent
in the one-loop level. For perturbation theory, we introduce the gauge-fixing
term and the Faddeev--Popov ghost term,
\begin{align}
   S_{\text{gf}}
   &=\frac{\lambda_0}{2g_0^2}
   \int d^Dx\,\partial_\mu A_\mu^a(x)\partial_\nu A_\nu^a(x),
\label{eq:(10)}
\\
   S_{c\Bar{c}}
   &=-\frac{1}{g_0^2}\int d^Dx\,
   \Bar{c}^a(x)\partial_\mu D_\mu ^{ab}c^b(x).
\label{eq:(11)}
\end{align}
We set $\lambda_0=1$ (the Feynman gauge) in what follows. These terms are not
invariant under the super transformation,
\begin{align}
   \delta_\xi S_{\text{gf}}
   &=-\int d^Dx\,\Bar{\xi}X_{\text{gf}}(x),&
   X_{\text{gf}}(x)
   &=\frac{\lambda_0}{g_0}
   \partial_\mu\partial_\nu A_\nu^a(x)\gamma_\mu\psi^a(x),
\label{eq:(12)}
\\
   \delta_\xi S_{c\Bar{c}}
   &=-\int d^Dx\,\Bar{\xi}X_{c\Bar{c}}(x),&
   X_{c\Bar{c}}(x)
   &=\frac{1}{g_0}f^{abc}\partial_\mu\Bar{c}^a(x)c^b(x)\gamma_\mu\psi^c(x).
\label{eq:(13)}
\end{align}
As noted in the footnote~\ref{foot-2}, $\delta_\xi S$ does not vanish
in~$D=4-2\epsilon$,
\begin{equation}
   \delta_\xi S
   =\int d^Dx\,
   \left[\partial_\mu\Bar{\xi}(x)s_\mu (x)
   -\Bar{\xi}(x)X_{\text{Fierz}}(x)\right],
\label{eq:(14)}
\end{equation}
where
\begin{equation}
   X_{\text{Fierz}}(x)
   =\frac{1}{2}g_0f^{def}\gamma_\mu
   \psi^d(x)\Bar{\psi}^e(x)\gamma_\mu\psi^f(x).
\label{eq:(15)}
\end{equation}

With the above three breaking terms, we have a WT relation,
\begin{align}
   &\left\langle
   \left[\partial_\mu s_\mu(x)
   +X_{\text{Fierz}}(x)+X_{\text{gf}}(x)+X_{c\Bar{c}}(x)\right]
   A_\alpha^b(y)\Bar{\psi}^c(z)\right\rangle
\notag
\\
   &=-\left\langle
   \delta(x-y)g_0\gamma_\alpha\psi^b(y)\Bar{\psi}^c(z)\right\rangle
   -\left\langle\delta(x-z)A_\alpha^b(y)
   \frac{1}{2g_0}\sigma_{\rho\sigma}F_{\rho\sigma}^c(z)
   \right\rangle.
\label{eq:(16)}
\end{align}
The effect of~$X_{\text{Fierz}}(x)$ can be taken into account as
\begin{align}
   &\left\langle
   \left[\partial_\mu s_\mu(x)
   +X_{\text{gf}}(x)+X_{c\Bar{c}}(x)\right]
   A_\alpha^b(y)\Bar{\psi}^c(z)\right\rangle'
\notag
\\
   &=-\left\langle
   \delta(x-y)g_0\gamma_\alpha\psi^b(y)\Bar{\psi}^c(z)\right\rangle'
   -\left\langle\delta(x-z)A_\alpha^b(y)
   \frac{1}{2g_0}\sigma_{\rho\sigma}F_{\rho\sigma}^c(z)
   \right\rangle',
\label{eq:(17)}
\end{align}
where the prime (${^\prime}$) implies that the expectation values are computed
with respect to the action with a
counterterm~$S'\equiv-\frac{1}{(4\pi)^2}C_2(G)\frac{1}{6}
\int d^Dx\,F_{\mu\nu}^a(x)F_{\mu\nu}^a(x)$. This follows from the fact that the
one-loop level expectation
value~$\langle X_{\text{Fierz}}(x)A_\alpha^b(y)\Bar{\psi}^c(z)\rangle$ can be
compensated by the variation of the
counterterm~$\langle\delta_\xi S'A_\alpha^b(y)\Bar{\psi}^c(z)\rangle$.

We want to find the correctly-normalized supercurrent which induces the
renormalized super transformation on \emph{renormalized\/} (elementary)
fields. For this, we first replace all the fields and couplings in the WT
relation~\eqref{eq:(17)} by renormalized ones. With the
notation~$\Delta\equiv\frac{g^2}{(4\pi)^2}C_2(G)\frac{1}{\epsilon}$, bare
fields/couplings and the renormalized ones are related in the one-loop level
as
\begin{align}
   g_0&=\mu^\epsilon\left(1-\frac{3}{2}\Delta\right)g,
\label{eq:(18)}
\\
   \lambda_0&=(1-\Delta)\lambda,
\label{eq:(19)}
\\
   A_\mu ^a(x)
   &=(1-\Delta)A_{\mu R}^a(x),\qquad
   \psi^a(x)=\left(1-\frac{1}{2}\Delta\right)\psi_R^a(x),
\label{eq:(20)}
\\
   c^a(x)&=\left(1-\frac{5}{4}\Delta\right)c_R^a(x),
\label{eq:(21)}
\\
   F_{\mu\nu}^a(x)
   &=\left(1-\frac{5}{2}\Delta\right)
   \left[\partial_\mu A_{\nu R}^a(x)-\partial_\nu A_{\mu R}^a(x)\right]
   +\left(1-\frac{11}{4}\Delta\right)f^{abc}\left[A_\mu^b(x)A_\nu^c(x)\right]_R.
\label{eq:(22)}
\end{align}
Substituting these into~Eq.~\eqref{eq:(17)} and evaluating UV divergences
coming from 1PI one-loop diagrams containing composite operators, after
rearrangements of various terms, we have~\cite{Hieda:2017sqq}
\begin{align}
   &\left\langle
   \left[\partial_\mu s_{\mu R}(x)
   +X_{\text{gf}R}(x)+X_{c\Bar{c}R}(x)
   \right]
   A_{\alpha R}^b(y)\Bar{\psi}_R^c(z)\right\rangle'
\notag
\\
   &=-\left\langle\delta(x-y)
   g\gamma_\alpha\psi_R^b(y)\Bar{\psi}_R^c(z)
   \right\rangle'
\notag
\\
   &\qquad{}
   -\left\langle\delta(x-z)A_{\alpha R}^b(y)
   \frac{1}{2g}\sigma_{\rho\sigma}
   \left[\partial_\rho A_{\sigma R}^c(z)
   -\partial_\sigma A_{\rho R}^c(z)
   +f^{cde}(A_\rho^d(x)A_\sigma^e(z))_R
   \right]\right\rangle'.
\label{eq:(23)}
\end{align}
For the definition of various renormalized composite operators,
see Ref.~\cite{Hieda:2017sqq}.

Equation~\eqref{eq:(23)} tells us that the finite
combination~$\partial_\mu s_{\mu R}(x)+X_{\text{gf}R}(x)+X_{c\Bar{c}R}(x)$ induces
the renormalized super transformation on renormalized fields in the one-loop
level. We can further show that the combination
$X_{\text{gf}}(x)+X_{c\Bar{c}}(x)$ vanishes in on-shell correlation functions
of gauge-invariant operators~\cite{Hieda:2017sqq}. Thus, in such correlation
functions, the correctly-normalized supercurrent to the one-loop level is given
by
\begin{equation}
   \mathcal{S}_{\mu R}(x)
   =s_{\mu R}(x)
   =-\frac{1}{2g_0}\sigma_{\rho\sigma}\gamma_\mu\psi^a(x)
   F_{\rho\sigma}^a(x)+\mathcal{O}(g_0^3).
\label{eq:(24)}
\end{equation}

\section{Supercurrent in terms of the flowed fields}
\label{sec-2}
In the previous section, we found that
$\mathcal{S}_{\mu R}(x)=%
-\frac{1}{2g_0}\sigma_{\rho\sigma}\gamma_\mu\psi^a(x)F_{\rho\sigma}^a(x)+%
\mathcal{O}(g_0^3)$ gives rise to the correctly-normalized supercurrent. We now
express this composite operator in terms of flowed fields $B_\mu (t,x)$
and~$\chi(t,x)$ for a small flow time.

We adopt the flow equations in~Ref.~\cite{Luscher:2013cpa}:
\begin{align}
   \partial_tB_\mu^a(t,x)
   &=D_\nu^{ab}G_{\nu\mu}^b(t,x),&
   B_\mu^a(t=0,x)&=A_\mu^a(x),
\label{eq:(25)}
\\
   \partial_t\chi^a(t,x)
   &=\left(D^2\right)^{ab}\chi^b(t,x),&
   \chi^a(t=0,x)&=\psi^a(x),
\label{eq:(26)}
\\
   \partial_t\Bar{\chi}^a(t,x)
   &=\Bar{\chi}^b(t,x)\left(\overleftarrow{D}^2\right)^{ba},&
   \Bar{\chi}^a(t=0,x)&=\Bar{\psi}^a(x),
\label{eq:(27)}
\end{align}
where
\begin{align}
   D_\mu^{ab}&=\delta^{ab}\partial_\mu+f^{acb}B_\mu^c(t,x),
\label{eq:(28)}
\\
   G_{\mu\nu}^a(x)&
   =\partial_\mu B_\nu^a(t,x)-\partial_\nu B_\mu^a(t,x)
   +f^{abc}B_\mu^b(t,x)B_\nu^c(t,x).
\label{eq:(29)}
\end{align}

Our goal is to rewrite the operator~$\psi^a(x)F_{\mu\nu}^a(x)$ in the
supercurrent~\eqref{eq:(24)} by the flowed fields. For this, we first expand
the flowed composite operator~$\chi^a(t,x)G_{\mu\nu}^a(t,x)$ in terms of
unflowed composite operators assuming that the flow time~$t$ small, i.e., we
consider a small flow-time expansion of~$\chi^a(t,x)G_{\mu\nu}^a(t,x)$. Noting
that the flow time~$t$ has mass-dimension~$-2$ and the flow equations are
Lorentz/gauge covariant, the small flow-time expansion takes the form
\begin{align}
   \chi^a(t,x)G_{\mu\nu}^a(t,x)
   &=\zeta_1(t)\psi^a(x)F_{\mu\nu}^a(x)
\notag
\\
   &\qquad{}
   +\zeta_2(t)
   \left[\gamma_\mu\gamma_\rho\psi^a(x)F_{\rho\nu}^a(x)
   -\gamma_\nu\gamma_\rho\psi^a(x)F_{\rho\mu}^a(x)\right]
\notag
\\
   &\qquad\qquad{}
   +\zeta_3(t)\sigma_{\rho\sigma}\sigma_{\mu\nu}\psi^a(x)F_{\rho\sigma}^a(x)
   +\mathcal{O}(t).
\label{eq:(30)}
\end{align}
We compute the three coefficients $\zeta_i(t)$ by perturbation theory;
perturbation theory is justified for the small flow-time limit $t\to0$ by the
asymptotic freedom. In~Eq.~\eqref{eq:(30)}, the off-diagonal operator mixings
arise only through loop corrections. Thus, to the one-loop order, we have
\begin{align}
   \psi^a(x)F_{\mu\nu}^a(x)
   &=\left[1-\zeta_1^{(1)}(t)\right]
   \chi^a(t,x)G_{\mu\nu}^a(t,x),
\notag
\\
   &\qquad{}
   +\zeta_2^{(1)}(t)\left[
   \gamma_\mu\gamma_\rho\chi^a(t,x)G_{\rho\nu}^a(t,x)
   -\gamma_\nu\gamma_\rho\chi^a(t,x)G_{\rho\mu}^a(t,x)\right]
\notag
\\
   &\qquad\qquad{}
   +\zeta_3^{(1)}(t)\sigma_{\rho\sigma}\sigma_{\mu\nu}
   \chi^a(t,x)G_{\rho\sigma}^a(t,x)+\mathcal{O}(t),
\label{eq:(31)}
\end{align}
where $\zeta_i^{(1)}$ are one-loop quantities. Substituting this into the
expression of the supercurrent~\eqref{eq:(24)}, we have
\begin{align}
   \mathcal{S}_{\mu R}(x)
   &=-\frac{1}{2g_0}\left[
   1-\zeta_1^{(1)}(t)-2(D-3)\zeta_2^{(1)}(t)+(D-9)(D-4)\zeta_3^{(1)}(t)
   \right]\sigma_{\rho\sigma}\gamma_\mu\chi^a(t,x)G_{\rho\sigma}^a(t,x)
\notag
\\
   &\qquad{}
   -\frac{1}{2g_0}\left[
   4(D-4)\zeta_2^{(1)}(t)-4(D-5)(D-4)\zeta_3^{(1)}(t)\right]
   \gamma_\rho\chi^a(t,x)G_{\rho\mu}^a(t,x)
\notag
\\
   &\qquad\qquad{}
   +\mathcal{O}(t)+\mathcal{O}(g_0^{3}).
\label{eq:(32)}
\end{align}

For the computation of the coefficients~$\zeta_i^{(1)}(t)$, it is convenient to
utilize the background field method~\cite{Suzuki:2015bqa}. We thus decompose
all the fields into the background fields (indicated by the hat (\,$\Hat{}$\,)) and
the quantum fields as
\begin{align}
   A_\mu^a(x)&=\Hat{A}_\mu^a(x)+a_\mu^a(x),&
   B_\mu^a(t,x)&=\Hat{B}_\mu(t,x)^a+b_\mu^a(t,x),
\label{eq:(33)}
\\
   \psi^a(x)&=\Hat{\psi}^a(x)+p^a(x),&
   \chi^a(t,x)&=\Hat{\chi}^a(t,x)+k^a(t,x),
\label{eq:(34)}
\\
   \Bar{\psi}^a(x)&=\Hat{\Bar{\psi}}^a(x)+\Bar{p}^a(x),&
   \Bar{\chi}^a(t,x)&=\Hat{\Bar{\chi}}^a(t,x)+\Bar{k}^a(t,x).
\label{eq:(35)}
\end{align}
With these decompositions, Eq.~\eqref{eq:(30)} becomes
\begin{align}
   &\left[\Hat{\chi}^a(t,x)+k^a(t,x)\right]
   \Bigl[
   \Hat{F}_{\mu\nu}^a(x)+\Hat{D}_\mu^{ab} b_\nu^b(t,x)-\Hat{D}_\nu^{ab} b_\mu^b(t,x)
   +f^{abc}b_\mu^b(t,x)b_\nu^c(t,x)\Bigr]
\notag
\\
   &\qquad{}
   -\left[\Hat{\psi}^a(x)+p^a(x)\right]
   \Bigl[
   \Hat{F}_{\mu\nu}^a(x)+\Hat{D}_\mu^{ab} a_\nu^b(x)-\Hat{D}_\nu^{ab} a_\mu^b(x)
   +f^{abc}a_\mu^b(x)a_\nu^c(x)\Bigr]
\notag
\\
   &=\zeta_1^{(1)}(t)\Hat{\psi}^a(x)\Hat{F}_{\mu\nu}^a(x)
   +\zeta_2^{(1)}(t)
   \left[\gamma_\mu\gamma_\rho\Hat{\psi}^a(x)\Hat{F}_{\rho\nu}^a(x)
   -\gamma_\nu\gamma_\rho\Hat{\psi}^a(x)\Hat{F}_{\rho\mu}^a(x)\right]
\notag
\\
   &\qquad{}
   +\zeta_3^{(1)}(t)
   \sigma_{\rho\sigma}\sigma_{\mu\nu}\Hat{\psi}^a(x)\Hat{F}_{\rho\sigma}^a(x)
   +\mathcal{O}(t).
\label{eq:(36)}
\end{align}
We compute the one-loop expectation value of the both sides by using
propagators of quantum fields in the presence of the background fields
(see~Ref.~\cite{Hieda:2016lly}). Some calculation gives us
\begin{align}
   \zeta_1^{(1)}(t)
   &=\frac{g_0^2}{(4\pi)^2}C_2(G)\frac{-2}{D-4}(8\pi t)^{2-D/2},
\label{eq:(37)}
\\
   \zeta_2^{(1)}(t)
   &=\frac{g_0^2}{(4\pi)^2}C_2(G)\frac{2}{(D-4)(D-2)}(8\pi t)^{2-D/2},
\label{eq:(38)}
\\
   \zeta_3^{(1)}(t)
   &=\frac{g_0^2}{(4\pi)^2}C_2(G)\frac{4}{(D-4)(D-2)}(8\pi t)^{2-D/2}.
\label{eq:(39)}
\end{align}
Equation~\eqref{eq:(30)} then yields
\begin{align}
   \mathcal{S}_{\mu R}(x)
   &=-\frac{1}{2g_0}
   \left[1+\frac{g_0^2}{(4\pi)^2}C_2(G)\frac{2(D-18)}{(D-2)D}(8\pi t)^{2-D/2}
   \right]
   \sigma_{\rho\sigma}\gamma_\mu\chi^a(t,x)G_{\rho\sigma}^a(t,x)
\notag
\\
   &\qquad{}
   -\frac{1}{2g_0}\frac{g_0^2}{(4\pi)^2}C_2(G)\frac{8(D-10)}{(D-2)D}(8\pi t)^{2-D/2}
   \gamma_\nu\chi^a(t,x)G_{\nu\mu}^a(t,x)+\mathcal{O}(t)+\mathcal{O}(g_0^3).
\label{eq:(40)}
\end{align}
We rewrite this expression by the renormalized gauge coupling~$g$ in the MS
scheme and the modified flowed gaugino
field~$\mathring{\chi}(t,x)$\footnote{Unlike the flowed gauge field, the flowed
fermion field requires the wave function renormalization. The modified flowed
gaugino field~$\mathring{\chi}(t,x)$ defined by
\begin{align}
   \mathring{\chi}(t,x)
   &=\sqrt{\frac{-\dim(G)}
   {(4\pi)^2t^2\langle\Bar{\chi}(t,x)\gamma_\mu(D_\mu-\overleftarrow{D}_\mu)
   \chi(t,x)\rangle}}
   \chi(t,x)
\notag
\\
   &=\frac{1}{(8\pi t)^{\epsilon/2}}
   \left\{1+\frac{g^2}{(4\pi)^2}C_2(G)
   \left[\frac{3}{2}\frac{1}{\epsilon}+\frac{3}{2}\ln{(8\pi\mu^2t)}
   -\frac{1}{2}\ln{(432)}\right]+\mathcal{O}(g^4)\right\}\chi(t,x)
\label{eq:(41)}
\end{align}
can avoid the explicit usage of the wave function renormalization constant.}
\begin{align}
   \mathcal{S}_{\mu R}(x)
   &=-\frac{1}{2g}
   \left[
   1+\frac{g^2}{(4\pi)^2}C_2(G)
   \left[-\frac{7}{2}-\frac{3}{2}\ln{(8\pi\mu^2t)}+\frac{1}{2}\ln{(432)}
   \right]\right\}
   \sigma_{\rho\sigma}\gamma_\mu\mathring{\chi}^a(t,x)G_{\rho\sigma}^a(t,x)
\notag
\\
   &\qquad{}
   -\frac{g}{(4\pi)^2}C_2(G)3\gamma_\nu\mathring{\chi}^a(t,x)G_{\nu\mu}^a(t,x)
   +\mathcal{O}(t)+\mathcal{O}(g^3).
\label{eq:(42)}
\end{align}
As Eq.~\eqref{eq:(24)} shows, the supercurrent does not depend on the
renormalization scale~$\mu$ when expressed by the running coupling constant.
We then may set the renormalization scale as~$\mu=1/\sqrt{8t}$ to yield
\begin{align}
   \mathcal{S}_{\mu R}(x)
   &=-\frac{1}{2\Bar{g}(1/\sqrt{8t})}
   \left\{1+\frac{\Bar{g}^2(1/\sqrt{8t})}{(4\pi)^2}C_2(G)
   \left[-\frac{7}{2}-\frac{3}{2}\ln\pi+\frac{1}{2}\ln{(432)}\right]
   \right\}
   \sigma_{\rho\sigma}\gamma_\mu\mathring{\chi}^a(t,x)G_{\rho\sigma}^a(t,x)
\notag
\\
   &\qquad{}
   -\frac{\Bar{g}(1/\sqrt{8t})}{(4\pi)^2}C_2(G)
   3\gamma_\nu\mathring{\chi}^a(t,x)G_{\nu\mu}^a(t,x)
   +\mathcal{O}(t)+\mathcal{O}(\Bar{g}^3(1/\sqrt{8t})).
\label{eq:(43)}
\end{align}
Finally, by taking the $t\to0$ limit, we obtain the desired expression,
\begin{align}
   &\mathcal{S}_{\mu R}(x)
\notag
\\
   &=\lim_{t\to0}
   \Biggl(-\frac{1}{2\Bar{g}(1/\sqrt{8t})}
   \left\{1+\frac{\Bar{g}^2(1/\sqrt{8t})}{(4\pi)^2}C_2(G)
   \left[-\frac{7}{2}-\frac{3}{2}\ln\pi+\frac{1}{2}\ln{(432)}\right]
   \right\}
   \sigma_{\rho\sigma}\gamma_\mu\mathring{\chi}^a(t,x)G_{\rho\sigma}^a(t,x)
\notag
\\
   &\qquad\qquad{}
   -\frac{\Bar{g}(1/\sqrt{8t})}{(4\pi)^2}
   C_2(G)3\gamma_\nu\mathring{\chi}^a(t,x)G_{\nu\mu}^a(t,x)
   \Biggr).
\label{eq:(44)}
\end{align}

\section{Conclusion}
\label{sec-3}
In this work, we constructed the correctly-normalized supercurrent in the 4D
$\mathcal{N}=1$ SYM in the Wess--Zumino gauge by using the gradient flow and
the small flow-time expansion. For this, we determined the renormalized
supercurrent in dimensional regularization in the one-loop level
(Sec.~\ref{sec-1}). Then we computed the small flow-time expansion of a
composite operator~\eqref{eq:(30)} (Sec.~\ref{sec-2}). The obtained
expression~\eqref{eq:(44)}, being UV finite, is independent of the
regularization method and hence in particular can be used with the lattice
regularization. We hope our representation of the supercurrent will be useful
in tuning the gaugino mass towards the supersymmetric point in the continuum
limit.\footnote{See Ref.~\cite{Kadoh:2016eju} for a recent review on
supersymmetric theories on the lattice.}

It is interesting to extend the present study to supersymmetric theories which
contain matter multiplets; for those theories, the parameter tuning to the
supersymmetric point will be quite demanding. In order to treat these extended
theories, we have to consider also the flow of the scalar field~$\phi(x)$. The
simplest choice of the flow equation would be
\begin{equation}
   \partial_t\varphi(t,x)=D_\mu D_\mu\varphi(t,x),\qquad
   \varphi(t=0,x)=\phi(x).
\label{eq:(45)}
\end{equation}
We are now studying the small flow-time representation of the supercurrent in
the 4D $\mathcal{N}=2$ SYM.

\section*{Acknowledgements}
We would like to thank Gernot M\"unster and Issaku Kanamori for fruitful
discussions during the conference.
This work was supported by JSPS Grant-in-Aid for Scientific Research Grant
Numbers JP16J02259 (A.~K.) and~JP16H03982 (H.~S.).

\end{document}